# A Thematic Study of Requirements Modeling and Analysis for Self-Adaptive Systems

Zhuoqun Yang
Institute of Mathematics, AMSS
Chinese Academy of Sciences
Haidian Dstr., Beijing
100190
zhuoqun.y@hotmail.com

Zhi Li
Software Engineering Dept.
Guangxi Normal University
Guilin, Guangxi
541004
zhili@gxnu.edu.cn

Zhi Jin
Key Laboratory of HCST (MoE)
Peking University
Haidian Dstr., Beijing
100187
zhijin@pku.edu.cn

## ABSTRACT
Over the last decade, researchers and engineers have developed a vast body of methodologies and technologies in requirements engineering for self-adaptive systems. Although existing studies have explored various aspects of this topic, few of them have categorized and summarized these areas of research in requirements modeling and analysis. This study aims to investigate the research themes based on the utilized modeling methods and RE activities. We conduct a thematic study in the systematic literature review. The results are derived by synthesizing the extracted data with statistical methods. This paper provides an updated review of the research literature, enabling researchers and practitioners to better understand the research themes in these areas and identify research gaps which need to be further studied.

## Categories and Subject Descriptors
D.3.1 [**Software Engineering**]: Requirements/Specifications – *Languages and Methodologies.*

## General Terms
Theory

## 1. INTRODUCTION
A self-adaptive system is a novel computing paradigm in which the system is capable of adjusting its behavior in response to meaningful changes in the environment and itself. Due to the inherent volatility of the deployed environment and frequent interactions with other software, systems, devices, sensors and people, self-adaptive systems (SASs) are facing the challenges brought by inadequate domain knowledge and incorrect domain assumptions. The adaptability is characterized by self-* properties [1]. When designing and developing SASs, engineers should take not only domain logic but adaptation logic into account.

Requirements engineering (RE) comes as the first stage in the development of software systems, aiming at defining domain logic, identifying stakeholders and their needs, and documenting them for subsequent analysis and implementation [2]. Different from RE for domain logic, RE for adaptation logic provides answers to what the adaptation requirements are, when and where the system needs to adjust themselves, who is responsible for achieving the adaptation, why the adaptation needs to be performed and how to achieve the adaptation.

During the RE process, requirements analysts should conduct several activities, e.g., requirements elicitation and subsequent requirements modeling and analysis. The modeling and analysis activities consist of requirements modeling, requirements specifying, system modeling, requirements validation and verification (V&V) [3]. Requirements modeling is the fundamental activity in the RE process. It abstractly describes what the target system should achieve by formulating the organization's structures, business rules, goals, tasks, behavior of stakeholders and systems, etc. Requirements models and system models can serve as the bases which are amenable to the following requirements activities, e.g., consistency checking, requirements verification.

Over the last decade, software researchers and engineers have developed a vast body of work on requirements modeling and analysis for SASs. Existing roadmaps and surveys [4-10] have summarized the methodologies and technologies proposed in requirements engineering for SASs, provided insights and outlined challenges for research work. However, to the best of our knowledge, no systematic study has been performed on summarizing the emerged research themes of the existing studies.

The overall objective of this review article is to systematically investigate the research literature of requirements modeling and analysis for self-adaptive systems from 2003 to 2013. We aim to summarize the state-of-the-art research themes generated from the involved modeling methods and requirements activities. This review presents significant extensions to our preliminary review of the literature in [11] with wider time span and more detailed results and discussions. To conduct the investigation and report systematic analysis results, we adopt the research guidelines of systematic literature review [12] in the evidence-based software engineering paradigm [13].

The rest of the article is structured as follows. Section 2 presents the brief of the research methodology and the review protocol, followed by the results of the extracted modeling methods, requirements activities and thematic study in Section 3. Section 4 provides discussions of threats to validity and corresponding countermeasures, followed by the related work in Section 5. Finally, Section 6 concludes the paper and suggests recommendations for future work.

## 2. RESEARCH METHODOLOGY
We brief the review protocol in this section. Readers can find more details of our protocol in [14].

### 2.1 Research Questions
We designed 3 research questions for this SLR:
**RQ1**: What modeling methods were utilized in relevant studies?

**RQ2**: What requirements activities were concerned?
**RQ3**: What research themes can be generated?
**RQ4**: How each theme was supported by the modeling methods?

## 2.2 Research Mechanism

### 2.2.1 Search Resources

To improve the quality of relevant studies, we choose 11 relevant conferences and symposiums (Table 1) and 10 relevant journals and books (Table 2) according to the Australian Research Council ERA (Excellence in Research for Australia) ranking[1]. To ensure thorough retrieval, we selected 8 common used digital libraries: IEEE Xplore, ACM Digital Library, Springer, Science Direct, Wiley InterScience, CiteSeerX, EI Compendex and Web of Knowledge.

**Table 1. Conferences and symposiums**

| Abbr. | Full name | Rank |
|---|---|---|
| ICSE | International Conference on Software Engineering | A |
| FSE | ACM SIGSOFT Symposium on the Foundation of Software Engineering | A |
| ASE | International Conference on Automated Software Engineering | A |
| RE | International Conference on Requirements Engineering | A |
| REFSQ | International Conference on Requirements Engineering: Foundation for Software Quality | B |
| MODELS | International Conference on Model Driven Engineering Languages and Systems | B |
| CAiSE | International Conference on Advanced Information Systems Engineering | B |
| ICAC | International Conference on Autonomic Computing | B |
| SASO | International Conference on Self-Adaptive and Self-Organizing Systems | N/A |
| SEAMS | International Symposium on Software Engineering for Adaptive and Self-Managing Systems | N/A |
| RE@runtime | International Workshop on Requirements @Run.Time | N/A |

**Table 2. Journals and books**

| Abbr. | Full name | Rank |
|---|---|---|
| TSE | IEEE Transactions on Software Engineering | A* |
| TOSEM | ACM Transactions on Software Engineering and Methodology | A* |
| IST | Information and Software Technology | A |
| JSS | Journal of Systems and Software | A |
| ESE | Empirical Software Engineering | A |
| ASEJ | Automated Software Engineering | A |
| REJ | Requirements Engineering Journal | A |
| SoSyM | Software and Systems Modeling | B |
| TAAS | ACM Transactions on Autonomous and Adaptive Systems | B |
| SESAS | Software Engineering for Self-Aaptive Systems | N/A |

### 2.2.2 Selection Strategy

Manual search is aimed to avoid missing important studies published in domain relevant venues. The retrieved results of automated search complement the results of manual search by expanding the coverage with more relevant studies. "Snowballing" search is another method for expanding the search coverage. For each search stage, we applied the selection criteria (Table 3). The relevant studies are derived with a 3-round selection: scan titles, read the abstracts and look through full-texts.

---

[1] Excellence in Research for Australia ranking (2010): Http://lamp.infosys.deakin.edu.au/era/?page=cnamesel10.

**Table 3. Selection criteria**

| ID | Inclusion criteria |
|---|---|
| C1 | Publication date between 2003.1-2013.12 |
| C2 | Peer-reviewed conference papers and journal articles |
| C3 | Involve concrete modeling methods and requirements activities |
| C4 | Involve concrete illustrations for the proposed approaches |
| **ID** | **Exclusion criteria** |
| C5 | Publication in non-English languages |
| C6 | Abstract, keynote, poster and short paper (less than 6 pages) |
| C7 | Opinion pieces and position papers |
| C8 | Secondary studies, e.g., roadmap, review and survey. |

### 2.2.3 Search and Selection Results

We scanned all papers in the publication venues with the selection criteria. The Kappa value was above 0.8 and any disagreement was eliminated by discussion. Finally, we selected 62 relevant papers in the end of the manual search (Table 4).

**Table 4. Frequency of papers in selected publication venues**

| Conference & symposium | Frequency | % | Journal & book | Frequency | % |
|---|---|---|---|---|---|
| SEAMS | 13 | 26% | REJ | 3 | 25% |
| RE | 6 | 12% | JSS | 3 | 25% |
| RE@runtime | 5 | 10% | SESAS | 2 | 17% |
| ICSE | 5 | 10% | IST | 1 | 8% |
| MODELS | 5 | 10% | TSE | 1 | 8% |
| REFSQ | 4 | 8% | ToSEM | 1 | 8% |
| ASE | 3 | 6% | ASEJ | 1 | 8% |
| CAiSE | 3 | 6% | ESE | 0 | 0% |
| FSE | 3 | 6% | SoSyM | 0 | 0% |
| ICAC | 2 | 4% | TAAS | 0 | 0% |
| SASO | 1 | 2% | — | — | — |
| Total | 50 | 100% | Total | 12 | 100% |

By analyzing the manual search results, we derived the search strings (Table 5) for automated search.

**Table 5. Search strings**

| ID | Search strings |
|---|---|
| S1 | **DOMAINS**[a] **AND** ("model requirements" **OR** "modeling requirements" **OR** "requirements modeling") |
| S2 | **DOMAINS AND** ("specify requirements" **OR** "specifying requirements" **OR** "requirements specifying" **OR** "requirements specification") |
| S3 | **DOMAINS AND** ("monitor requirements" **OR** "monitoring requirements" **OR** "requirements monitoring") |
| S4 | **DOMAINS AND** ("aware requirements" **OR** "requirements-aware" **OR** "requirements awareness" **OR** "requirements-awareness") |
| S5 | **DOMAINS AND** ("diagnose requirements" **OR** "diagnosing requirements" **OR** "requirements diagnosing" **OR** "requirements diagnosis") |
| S6 | **DOMAINS AND** ("detect requirements" **OR** "detecting requirements" **OR** "requirements detection") |
| S7 | **DOMAINS AND** ("verify requirements" **OR** "verifying requirements" **OR** "requirements verifying" **OR** "requirements verification") |
| S8 | **DOMAINS AND** requirements **AND** (adaptation **OR** reconfiguration **OR** decision) |
| S9 | **DOMAINS AND** ("evolution requirements" **OR** "requirements evolution") |

[a] **DOMAINS** denotes the string: ("self-adaptive systems" **OR** "dynamically adaptive systems" **OR** "self-adaptive software" **OR** "autonomic computing").

After automated search and "snowballing" search, we totally selected 109 papers which are provided in Appendix A.

## 2.3 Data Extraction and Thematic Synthesis

We mainly extract the modeling methods and requirements activities from primary studies. To generate the themes of the primary studies, we adopted thematic synthesis, which is a systematic and relatively objective method for identifying, integrating and reporting the themes or patterns based on raw data. It minimally organ-izes and describes the data set in rich detail and frequently interprets various aspects of the research themes [15]. We integrated the recommended steps in [15] and divided the synthesis process into *extracting textual segments*, *coding texts* and *translating codes into themes*

The coding process was performed by the two PhD students. Once all segments were coded, *basic themes* were derived by assembling studies with the same or similar codes together. Then, basic themes were aggregated into *organizing themes*. Finally, *global themes* were generated by comparing and integrating organizing themes. More details can be found in our protocol [14]. The details of extracted data can be found in [15].

## 3. RESULTS AND DISCUSSIONS

*RQ1:. What modeling methods were utilized in relevant studies?*

Totally, we extracted 29 modeling methods (Figure 1) from the primary studies, 34% (37/109) of which utilized more than one method while 66% (72/109) of which only utilized one method.

KAOS is the most widely used goal-oriented method in the literature (among KAOS, i* and Tropos). DTMC and MDP are most frequently utilized Kripke structures. Logic models can be used for activities at both requirements time and runtime. LTL and FL are most considered logic models. Except for enterprise models, behavior models and logic models, some models of Artificial Intelligence field are also adopted to model requirements, such as GA and GP. Utility can be used to representing the satisfaction of requirements.

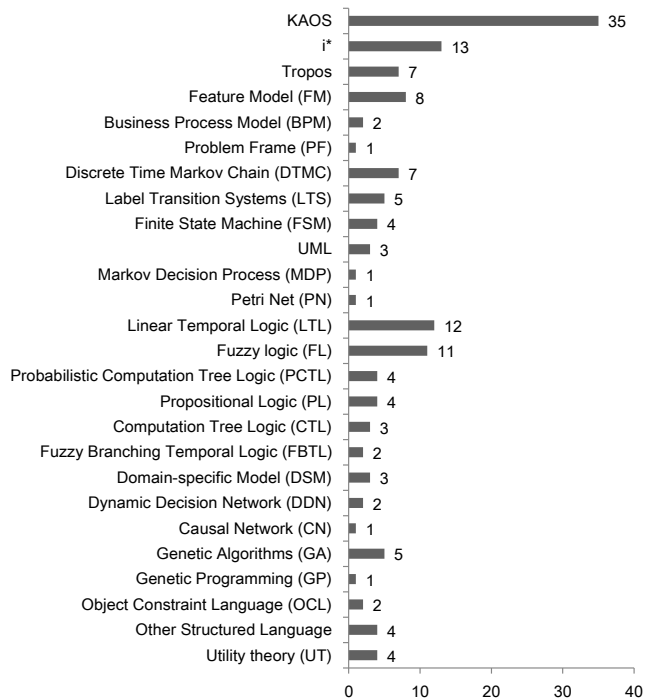

**Figure 1. Frequency of modeling methods.**

*RQ2: What requirements activities were concerned?*

The involved requirements activities in the primary studies are presented in Figure 2. The first six activities are performed at requirements time (44%). The last four activities are conducted at runtime (46%). The rest four belong to design and development time (10%).

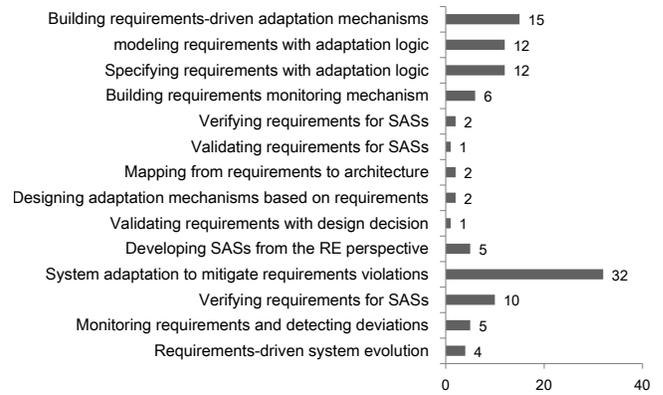

**Figure 2. Frequency of requirements activities.**

*RQ3: What research themes can be generated?*

The refinement of requirements activities can be achieved through thematic synthesis. We coded the text of the primary studies and aggregated similar codes into 35 basic themes. 12 organizing themes are generated from basic themes by referring to the requirements activities while 3 global themes are identified according to requirements activity timelines. The results of thematic synthesis and corresponding modeling methods and primary studies are presented in Table 6.

*RQ4: How each theme was supported by the modeling methods?*

*T1.1 Support for effective adaptation mechanisms*

The first basic theme concentrates on certain kinds of application systems, such as web service and Internetware. Baresi et al. [S1, S2, S7] conducted a series of work on building adaptation mechanisms for service systems based on KAOS models and FL. The original idea can be found in [S9]. Adaptive services can adjust their behavior and satisfy resilient requirements through adding, removing and modify services. Besides, the specifications of goals are derived by extending LTL with fuzzy notations. This kind of mechanism endows adaptive service with the quality attribute of fault-tolerance. The technique of service composition provides avenues for reconfiguration of adaptive services at runtime. Ma et al. [S6] proposed an adaptive mechanism for service systems through requirements monitoring and service composition. Other research work involved in this theme focuses on Internetware, which is a kind of software constructed by a set of autonomous software entities distributed over the Internet. The characteristic of structure claims modeling methods that can represent the distributed entities and their relationships. As described in Section 4.2.1, i* is appropriate for modeling this kind of system. Each part of Internetware is modeled as an actor. The connections between distributed entities are represented as kinds of dependencies. Jian et al. [S3, S4] proposed that the adaptation is stimulated by some triggers and achieved by selecting different tasks. Compared with them, Yang and Jin [S5] introduced an *Event-Condition-Action* (ECA) rules based parametric adaptation mechanism. Studies in this basic theme have two things in common: 1) systems are represented with enterprise models, especially goal-oriented models and 2) the adaptation mechanisms imply the processes of requirements monitoring, deviation diagnosing and decision making.

The second basic theme is generated according to special concerns during the adaptation process. Due to the inherent volatility of the environment and imperfect domain knowledge, we cannot predict all the changes at runtime. These unanticipated changes in both context and system can be considered as *uncertainties*. Thus, when building adaptation mechanisms, one special concern is how to deal with these uncertainties. Baresi et al. [S9] proposed FLAGS and applied fuzzy theory and membership functions to turn the crisp satisfaction of requirements into fuzzy satisfaction. When context changes unexpectedly, system can be still in operation with relatively low performances. Goldsby and Cheng [S8] addressed how to tackle uncertainty during development and how to generate behavior models for different environmental conditions. Another special concern is self-explanation. SASs need the capability of self-explanation to explain system's behavior during maintenance and garner confidence in customers [S10]. This concern is realized by adding some *claims* as notations to NFR modeled with i* framework. It is easy to find that these special concerns are closely related to software quality attributes: dealing with uncertainty provides fault-tolerance while self-explanation provides understandability.

The third basic theme emphasizes constructing adaptation mechanisms through various kinds of RE perspective. Ali et al. [S11] addressed the adaptation mechanism from the social aspect. Social adaptation refers to the adaptation which responds to social collective judgment about the correctness and efficiency of a system in meeting users' requirements. Social feedback becomes a primitive driver for adaptation and user perception becomes essential criteria for selecting alternative behavior. Santos et al. [S12] introduced how to represent variability in business process and select configurations. They considered context as adaptation trigger and NFR as criteria for selecting the reconfigured process. Salifu et al. [S13] described an adaptation mechanism from the perspective of problem frame. The context sensor is modeled as machine domain for the monitoring problem while the adaptation controller is modeled as machine domain for the switching problem. The adaptation problem is specified with statechart. Shen et al. [S14] proposed an adaptation mechanism from the software product line (SPL) perspective. Feature-oriented methods are used to represent alternative configurations. Quershi el al. [S15] considered adaptation mechanism from a more general view.

*T1.2 Modeling requirements and related concerns*

The first basic theme is related to the requirements modeling for certain types of system. Ahmad and Bruel [S16] discussed how to combine KAOS, SysML and RELAX in modeling requirements for Ambient Assisted Living systems. Sawyer el al. [S17] proposed three levels of analyzing dynamically adaptive systems (DASs), including monitoring, decision making and adaptation and provided i* based requirements model for each level. The requirements model can be translated into DSL for architecting middleware component configurations. Based on KAOS, Sutcliffe and Sawyer [S18] modeled requirements of personalized adaptive systems, which involve individual users in the adaptation loop.

**Table 6. Levels of themes and related modeling methods and studies.**

| Global theme | Organizing theme | Basic theme | Modeling method and primary study | # |
|---|---|---|---|---|
| T1. RE support at requirements phase | T1.1. Support for effective adaptation mechanisms | Adaptation mechanisms for specific adaptive systems | KAOS & FL [S1, S2, S7], i* [S3-S6] | 7 |
| | | Quality attributes supported by adaptation mechanisms | UML [S8], KAOS & FL [S9], i* [S10] | 3 |
| | | RE perspectives for building adaptation mechanisms | KAOS [S11, S15], BPM [S12], PF [S13], FM [14] | 5 |
| | T1.2. Modeling requirements and related concerns | Requirements models for specific adaptive systems | KAOS [S16, S18], i* [S17, S20], Tropos [S19] | 5 |
| | | Requirements models within changed environments | i* [S21], KAOS [S22] | 2 |
| | | Advantages supported by requirements models | KAOS [S23], i* [S24] | 2 |
| | | Empirical study on requirements models | Tropos [S25] | 1 |
| | | Modeling adaptation logic for design and development | DSM [S26, S27] | 2 |
| | T1.3. Specifying adaptation behavior and adaptation logic | Specifying adaptation behavior | LTS & LTL [S28, S32, S33], PN [S29], LTL [S30, S31] | 6 |
| | | Addressing uncertainty with specifications | SL [S34], FBTL & FL [S35, S36], FL [S37] | 4 |
| | | Specifying goal-based adaptation logic | KAOS, LTL [S38], SL [S39] | 2 |
| | T1.4. Support for effective requirements monitoring mechanisms | Monitoring mechanisms with context-aware ability | Tropos [S40], i* [S41] | 2 |
| | | Advantages supported by requirements monitoring | KAOS [S42] | 1 |
| | | Monitoring mechanisms built on requirements models | Tropos, PL [S43], KAOS & OCL [S44, S45] | 3 |
| | T1.5. Requirements verification | Verifying adaptation logic | LTS & LTL [S46], LTL [S47] | 1 |
| | T1.6. Requirements validation | Detecting inconsistency of conflicts | Tropos [S48] | 1 |
| T2. RE support at design and development phase | T2.1. Support for effective design decisions | Mapping from requirements to architecture | i* [S49], KAOS [S50] | 2 |
| | | Support for designing adaptation mechanisms | KAOS [S51], DSM [S52] | 2 |
| | | Requirements validation at design time | SL [S53] | 1 |
| | T2.2. Support for effective development | Development advanced by RE methods | DTMC & PCTL [S54], Tropos [S55], KAOS [S56] | 3 |
| | | Customized development of SASs | KAOS [S57], FM [S58] | 2 |
| T3. RE support at deployment phase (runtime) | T3.1. Adaptation and decision making at runtime | Requirements-driven adaptation for specific domain | Tropos [S59], KAOS [S60], KAOS & UML [S61] | 3 |
| | | Requirements-driven architectural adaptation | KAOS [S62, S63] | 2 |
| | | Control-based runtime adaptation | DTMC [S64, S65], KAOS [S66] | 3 |
| | | Runtime adaptation advanced by various models | FM [S67, S72], GA [S69, S72], PL [S69], KAOS [S70, S71, S73], FL [S74], BPM [S75] | 9 |
| | | NFR trade-off during adaptation | FM [S76], FSM [S77], GP [S78], GA [S79] | 4 |
| | | Coping with uncertainty during adaptation | FM [S80, S81], KAOS, UT & GA [S82], DDN & UT [S83], DDN & i* [S84], UML & MDP [S85], i* & FL [S86], FL [S87] | 8 |
| | | Adaptation and individual concerns | KAOS & CN [S88], KAOS [S89], LTS & LTL [S90] | 3 |
| | T3.2. Verifying requirements at runtime | Verifying adaptation logic | LTL [S91, S92], FSM & CLT [S93, S94] | 4 |
| | | Verifying requirements with probabilistic models | DTMC & PCTL [S95, S96, S97] | 3 |
| | | Verification with models@runtime | DTMC [S98] | 1 |
| | | Specifying verification process | FSM & CLT [S99], LTL [S100] | 2 |
| | T3.3. Monitoring requirements at runtime | Monitoring requirements and detecting violations | FL [S101], KAOS & PL [S102, S103], KAOS & UT [S104] | 4 |
| | | Optimizing monitoring configuration | GA & UT [S105] | 1 |
| | T3.4. System evolution at runtime | System evolution based on requirements evolution | KAOS [S106, S107], KAOS & FM [S108], SL [S109] | 4 |

This approach addresses the importance of user preferences and how to integrate the preference into the adaptation process. Morandini et al. [S19] adopted Trops4AS [S55] to deal with the requirements modeling of self-organizing multi-agent systems. Compared with these studies, Qureshi et al. [S20] brought forward a more general requirements modeling language based on Techne and i* framework, named Adaptive RML. Modeling methods used in this basic theme are all goal-oriented.

As discussed above, environmental changes are essential causes of adaptation. Some environmental changes are predictable while the others are not. The second basic theme aims to identify how to address sources of environmental changes, both certain ones and uncertain ones, during modeling requirements. Goldsby et al. [S21] proposed *LoREM*, i.e., Levels of RE for Modeling, to describe the modeling work performed by system developers. LoREM characterizes four levels of RE work, each of which should be done by different system developer. Developers identify several environmental changes and model the predictable information within monitoring tasks. Adaptation will be triggered according to the given thresholds of monitored information. However, when environmental changes are unanticipated, developers cannot specify all these information into the requirements model. Cheng et al. [S22] proposed an incremental modeling method to cope with environmental uncertainties. Potential threats to the target system are incrementally identified and linked to KAOS model elements. The identified uncertainties can be mitigated through relaxing goals or adding new goals.

The third basic theme focuses on some special concerns during modeling requirements. Control theory and feedback control have been successfully used in the RE for SASs literature [7]. In control theory, system identification is the process of determining the equations governing the system behavior. Once finding out the mathematical relationships between system inputs and outputs, we can build more efficient parametric adaptation mechanisms. Nevertheless, before software systems are implemented, i.e., at requirements time, they are gray box systems or black box systems, which present no explicit quantitative relationships. Souza et al. [S23] investigated the qualitative differential relations between configuration parameters and target outputs from RE perspective and proposed a reconciliation process based on the identified relations. Requirements management helps deal with the anticipated changes of requirements by recording the relationships between requirements and down-stream artifacts of the development process. Due to the inherently susceptibility of DASs, building traceability in requirements models is significant for providing evidence for requirements evolution. Welsh and Sawyer [S24] classified types of changes that have impacts on requirements and extended i* with the notion of *claims*. These notions help explain requirements changes.

Empirical study on requirements models concerned the research work aiming to evaluate requirements modeling methods through observation, experiment and answering questionnaires. Morandini et al. [S25] conducted a controlled experiment on evaluating the comprehension of requirements represented with Tropos4AS.

The fifth basic theme concerns how adaptation logic built at requirements phase can benefit the design and development phase. Luckey et al. [S26] discussed the relationship between the logical design phase and the technical design phase. They addressed that when modeling requirements with use cases, function aspects and logic aspects are always mixed in the same model. Having a dedicated modeling language for adaptivity aspects would support better software quality. They introduced adapt cases derived from traditional use cases and demonstrated how adapt cases can fill the gap between logical analysis and technical design. Based on the idea of separating functional concerns and adaptation concerns, Schilit and Theimer [S27] proposed a modeling language for specifying SASs based on adapt cases and described the development process supported by the modeling language.

### *T1.3 Specifying adaptation behavior and adaptation logic*

Specifying adaptation behavior focuses on modeling system behavior and provides logic-based specifications of requirements or properties that should be held by the behavior model. In this basic theme, several behavior models and logic models are utilized. Zhang and Cheng [S30] proposed a program model with a state machine and described semantics of three kinds of program adaptation. To specify the adaptation behavior, they extended LTL with new operators. They elaborated the proofs of safety properties and liveness properties of the adaptive program introduced in [S31]. Based on these ideas, the same authors [S29] introduced how to separate adaptation behavior from non-adaptation behavior of adaptive programs, and formalized the adaptive program and adaptation behavior with Petri Nets and LTL respectively. Inspired by their work, Zhao et al. [S28] developed specifications of reachability, safety, maintainability and integrity properties for adaptive programs and verified these properties with LTSA. Furthermore, D'ippolito et al. [S32, S33] presented the technique for synthesizing behavior models that works for an expressive subset of liveness properties.

To address uncertainties with specifications, the community is prone to utilize fuzzy theory. Whittle et al. led a series of studies [S34-S36] to produce RELAX, which is flexible specification language for representing both requirements and uncertainties. RELAX specifications are structured with lots of natural language and Boolean operators, such as *SHALL*, *AS EARLY AS POSSIBLE* and *BEFORE*. The semantics of RELAX is defined with FBTL, which describes a branching temporal model with fuzzy logic. Thus, the specifications also provide the property of fault-tolerance. Different from RELAX, Torres et al. [S37] addressed that the existing specification models using specific numbers to specific non-functional constraints may become obsolete at runtime. To cope with this kind of uncertainty, the constraints should not be assigned with static values.

Specifying goal-oriented adaptation logic concentrates on the specifications of goal models. Brown et al. [S38] presented the goal model of adaptive programs introduced in [S30] and incorporated the A-LTL specifications into the KAOS specifications. Sabatucci et al. [S39] proposed GoalSPEC, a goal-oriented specification language, to decouple business goals and their implementation. In this way, it may be flexible to derive plan compositions for satisfying goals.

### *T1.4 Support for effective requirements monitoring mechanisms*

The first basic theme claims the importance of monitoring context. Context can be defined as the reification of the environment [17]. The notion of context-aware was introduced in Section 2.2. To provide context-aware ability, how to represent contextual information should be addressed firstly. Ali et al. [18, 19] proposed a hierarchical context model for modeling contextual information and a contextual goal model for building the relationships between contexts and requirements. Based on these work, they provided an approach to optimizing the context that should be monitored [S40]. During operation, a set of tasks is selected to achieve the requirements, so the system only needs to monitor a reduced set of contextual information. In this way, the monitoring overhead can be reduced. Welsh and Sawyer [S41] addressed that a DAS monitors its environment and uses the monitored data to make adaptation decisions. However, when the environment is poorly understood, the decision making process may be limited by the imperfect domain assumptions. Therefore, to represent the environmental uncertainties, they added some *claims* to the requirements model. Once the *claims* are monitored and proven to be false, relevant solutions will be selected.

Advantages supported by requirements monitoring concerns the quality attributes assured with monitoring mechanisms. Pasquale et al. [S42] proposed a monitoring mechanism for engineering adaptive security. Attackers can use system vulnerability (secondary asset) to attack system and harm users' requirements (primary asset). Therefore, both the requirements and the vulnerability should be protected. They modeled vulnerability and countermeasures by adding notations to the goal model. Besides, a model of monitoring functions is built for updating assets.

The third theme focuses on how to monitor requirements captured within goal models. Dalpiaz et al. [S43] proposed a monitoring and diagnosis mechanism and related algorithms for identifying failures in goals and tasks which are modeled with Tropos method. This mechanism can also be applied to monitor requirements modeled by KAOS or i*. Souza et al. [S44, S45] provided a requirements monitoring framework by introducing *awareness requirements* (*AwReqs*). *AwReqs* is defined as requirements that refer to other requirements or domain assumptions and their success or failure. To help design and implementation, they provided the formalization of *AwReqs* with extended OCL.

*T1.5 Requirements verification*

The basic theme of this organizing theme is *Verifying adaptation logic*. Abeywickrama and Zambonelli [S46] introduced SOTA, i.e., State of the Affairs, as a general goal-oriented modeling framework for the analysis and design of self-adaptive systems. The goal model is transformed into a label transition system and requirements are specified with Fluent Linear Temporal Logic (FLTL). The verification process is performed through model checking with LTSA. This approach helps iteratively refine the requirements model and produce correct specifications. The verification work of Zhang et al. [S47] is the follow-up work of [S29-S31]. Steady-state programs are modeled as states in a behavior model while adaptation processes are modeled as transitions between states. Steady-state programs are specified with LTL while properties during adaptation are specified with A-LTL. Three modular algorithms are defined for verifying 1) properties of steady-state programs; 2) global invariants and 3) transitional properties when the programs adapt.

*T1.6 Requirements validation*

The selected studies of this organizing theme focus on Detecting inconsistency and conflicts. Ali et al. [S48] addressed how to identify and eliminate the inconsistent contexts and conflicting context based on the contextual goal model. Conflicting context implies that at certain context, two or more system actions lead to requirements conflicts. By detecting inconsistency, system requirements can be achieved with sets of alternative tasks. By detecting conflicts, it ensures that there is at least a conflict-free set of tasks. The authors provided a CASE tool and processes to guide the construction and analysis of contextual goal model.

*T2.1 Support for effective design decisions*

The first basic theme concerns how to derive architectural models from requirements models for SASs. Pimentel et al. [S49] proposed a systematic approach of incremental generating architectural models from requirements models. The requirements are modeled with i* framework. The generating process consists of six activities: *requirements refactoring* (modularizing the system), *context annotation and analysis* (defining related context), *identification of sensors and actuators* (defining interfaces), *generate architectural model* (converting models), *refine architectural model* (defining sub-component) and *include the self-adaptation component* (defining components that will perform reasoning). Tang et al. [S50] provided two views of software architecture: the structural view and the behavior view. The structural view is generated by transforming the KAOS model. The behavior view is defined with FSM and the behavior is formalized with Communicating Sequential Process (CSP). These studies introduce how to use goal models as a foundation to design a SAS. Besides, the transformation process provides the traceability and consistency between requirements models and architectural models.

The second basic theme addresses how requirements engineering can help design adaptive mechanisms. Souza and Mylopoulos [S51] reported on the application of *Zanshin* for the design of ambulance dispatch system. The *Zanshin* method focuses on three phases. The first phase, designing an adaptive system, can be divided in to three steps: eliciting *AwReqs*, system identification and specifying evolution requirements (EvoReqs). The second phase is integrating implementation with *Zanshin* framework in order to provide a generic feedback loop of adaptation based on the models produced earlier. The last phase is adding failures that trigger adaptation scenarios, in order to verify that the method can achieve requirements-driven adaptation. Souza and Mylopoulos concentrate on self-awareness, while Bencomo et al. [S52] proposed an approach to supporting the design and operation of DASs from the perspective of context-awareness. The approach models two variability dimensions: environment variability and structural variability. The former one defines the conditions under which the adaptation is needed, while the latter one defines alternative architectural configurations. The variability of the environment and structural configurations is modeled in terms of variability modeling from SPL. This work addressed environment variability offers a way to reason about environmental changes at runtime and functions as the trigger of system adaptation.

To support requirements validation at design time, Becker et al. [S53] applied RELAX to formal specifying NFR of SASs and proposed a model-based approach to semi-automatically analyzing whether the self-adaptation layer ensures the satisfaction of NFR. This approach provides a view of analyzing adaptation requirements at design time.

*T2.2 Support for effective development*

Development advanced by RE methods concentrates on how RE methods can benefit the implementation and development of SASs. Ghezzi and Sharifloo [S54] discussed how to ensure satisfaction of NFR at runtime by proposing a development process based on dynamic software product line (DSPL). When modeling DSPL, variation points and variants are specified with feature models. The varying behavior can be specified with sequence diagrams. To verify NFR at runtime, the sequence diagrams are transformed into DTMC and CTMC for performing model checking. When requirements are violated, new configurations can be realized through selecting candidate features. Morandini et al. [S55] provided a goal-oriented development process by identify-ing the goal model, the environment model and the fault model, and introducing how to implement these models in JADEX. Nakagawa et al. [S56] focused on the development process for supporting dynamic evolution of SASs. They introduced a development style for dynamically evolving systems by adding or replacing control loops that constitute the systems; devised a goal model compiler to generate models for evolution; and presented a programming framework that allows control loops to be changed at runtime.

Customization is another important concern during the development of SASs, because users expect the system to perform different behaviors according to individual needs. Since goal models and feature models both involve the notion of variation points and variants, they can both serve as models for realizing customization. Liaskos et al. [S57] applied goal models to represent alternative behaviors that users can exhibit to achieve their goals. Customization information is added to restrict the space of possibilities to those that fit specific users, contexts or situations. Elements of the goal model are mapped to units of source code. This way, customization preferences posed at the requirements level are directly translated into system customizations. Based on SPL and feature models, Shen et al. [S58] proposed a domain requirements model including business requirements, context and adaptation logic. The model contains a set of variability binding constraints that ensure the customization.

***T3.1*** *Adaptation and decision making at runtime*

Requirements-driven adaptation for specific domain concerns adaptation of certain kinds of adaptive systems. As the follow-up of [S43], Dalpiaz et al. [S59] proposed a requirements-driven architecture for adaptation of social-technical systems (STS). The architecture performs a monitor-diagnose-reconcile-compensate cycle, which is similar to the MAPE process. Requirements of STS are modeled with Tropos method. The adaptation cycle 1) monitors actors' behavior and contextual changes; 2) diagnoses failures and under-performance by checking whether monitored behavior is compliant with actors' goals; 3) finds a possible way to address the problem; and 4) enacts compensation actions to reconcile actual and desired behavior. Differently, Fu et al. [S61] presented a self-repairing approach for STS based on monitoring states of requirements. System requirements are modeled with Tropos methods. Each requirement is represented as a state machine. Repair decisions are made by maintaining each state machine. Chen et al. [S60] propose a requirements-driven self-tuning approach for the survivability assurance of web systems. Their approach is based on PID (proportional-integral-derivative) controller and goal-oriented reasoner for both FR and NFR. The goal-based configuration plans are produced by reasoning on the live goal model, and then mapped into system architectural configurations.

Different from the former basic theme, requirements-driven architectural adaptation focuses on domain-free adaptation approaches. Following [S56], Nakagawa et al. [S62] exploited the structure of the goal-oriented requirements description with constraints and a control loop pattern. They also provided an architectural compiler for self-adaptive systems, which generates architectural configurations from the goal-oriented requirements descriptions. Shen et al. [S63] proposed a quality-driven self-adaptation approach, which incorporates requirements-level and architecture-level adaptation. At the requirements level, value-based quality trade-off decisions are made. At the architecture level, component-based architecture adaptation is carried out. To bridge the gap between requirements and runtime architecture, design decisions capture alternative design options and help map requirements adaptation and contextual changes to architectural adaptation operations.

Control-based runtime adaptation is concerned with realizing adaptation with control-based approaches, particularly with feedback control. Filieri et al. [S64] proposed a reliability requirements-driven adaptation approach by dynamic binding components. The dynamic binding problem is formulated as a feedback control problem and solved with linear blocks. The same authors in [S65] provided an approach to ensuring reliability requirements at runtime through automatically modifying the model based on feedback control. The system is modeled as a DTMC. The transitions of a state represent alternative choices, which are assigned with probabilities. Reliability requirements are expressed as reachability properties that constrain the probability to reach certain states, denoted as failure states. The probabilities are modified at runtime by a feedback controller to ensure the satisfaction of the requirements. Peng et al. [S66] applied feedback control to tuning the preference of quality requirements. Optimal configurations can be derived through reasoning with goal models and making trade-off decisions with the tuned preferences.

Runtime adaptation advanced by various models addresses novel adaptation approaches based on various kinds of models. Several approaches benefit from the idea of models@runtime, which seeks to extend the applicability of models and abstractions to the runtime environment, with the goal of providing effective technologies for managing the complexity of evolving software behavior while it is executing. Moisan et al. [S67] proposed an approach for adaptation of large scale systems. Two feature models are produced by separating the concern of task requirements and system components. Intra- and cross- constraints are added to the two models. Runtime configurations can be derived by selecting features and model-based transformations. Torres et al. [S74] provided an approach to updating the specifications of quality of services requirements for service-based systems at runtime. When the updated specifications are not satisfied by the current configuration, an adaptation is triggered. Mosincat et al. [S75] considered system models as runtime entities. They presented how to support adaptability with evolved system models derived by integrating new variants and validating existing variants. Ramirez et al. [S68, S79] and Pascual et al. [S72] applied genetic algorithms to derive optimal adaptation decisions at runtime. The former study modeled variability with feature models and features are represented as genes for crossover and mutation. The latter used genes to represent the topology vector of a network. Some approaches involved in this basic theme are based on goal models. Salehie and Tahvildari [S70] proposed the goal-action-attribute model (GAAM) for modeling requirements and making adaptation decisions. To evaluate the action selection mechanism, they systematically designed a series of research questions and comparative experiments. Souza et al. [S71] introduced the *Qualia* framework that provides qualitative adaptation based on *AwReqs* and *Zanshin* framework. Wang and Mylopoulos [S73] proposed an autonomic architecture consisting of monitoring, diagnosis, reconfiguration and execution components. New configurations can be derived through performing monitoring, diagnosis and reasoning with KAOS requirements models. Besides, problem framework is also considered in this basic theme. As the extension of [S13], Salifu et al. [S69] introduced an approach of investigating the necessary and sufficient conditions for both monitoring and switching problems. These problems are encoded into propositional logic constraints and analyzed automatically using SAT solver.

The basic theme, NFR trade-off during adaptation, identifies the relevant studies which discuss NFR trade-off during decision making process. Horikoshi et al. [S76] leveraged feature models to identify adaptation point and NFR, and applied an architectural description language (ADL) to specifying components. At runtime, feature models and ADL descriptions are processed with a parser and new configurations are generated according to component dependencies between features and components. Then, NFR trade-off is achieved by calculating the contribution of newly selected features to NFR. Mori et al. [S77] focused on the trade-off between user preference and reconfiguration costs. They proposed a multi-objective utility function to determine the best reconfiguration. Ramirez et al. [S78] proposed an evolutionary computation based approach to evolving adaptation paths while balancing the trade-off between FR and NFR.

Uncertainties may occur both inside the systems and within the environment. First, it is often infeasible to precisely detect, measure and describe all the contextual changes. This imprecision in the context is viewed as the environmental or external uncertainty. On the other hand, system or internal uncertainty contains various kinds, such as sensor errors or failures, satisfaction of NFR and side effects of adaptation. Therefore, dealing with these two kinds of uncertainties is a challenging theme in the literature. Elkhodary et al. [S80, S81] presented the FUSION approach which uses machine learning methods to tune the adaptive behavior of the system to unanticipated changes. Esfahani et al. [S87] proposed the POISED approach which aims to improve the quality attributes of a software system and achieve a global optimal configuration through reconfiguration of components. Ramirez et al. [S82] introduced the AutoRELAX approach and applied the genetic algorithm to automatically generating RELAXed goal model for addressing environmental uncertainty. By searching for the goal model, a DAS can derive its adaptation decisions. Ramirez et al. [S86] integrated RELAX language and the notation of *Claims* in order to assess the validity of *Claims* at run time while tolerating minor and unanticipated environmental conditions that can trigger adaptation. Bencomo and Belaggoun [S84] proposed to use DDN to support the decision making process with environmental uncertainty. Tasks, softgoals and contributions of tasks to softgoals are mapped to decision nodes, chance nodes and utility correspondingly. The adaptation decisions are derived through reasoning with probabilities. A well-conducted case study for presenting how the probabilistic model can be used to support the decision-making in SASs is provided in [S83] by the same authors. Moreover, Ghezzi et al. [S85] concentrated on dealing with non-functional uncertainties, i.e., response time and faulty behavior of components. The system is modeled as a finite state automaton augmented with probabilities and is adapted by choosing among alternative possible paths of the automaton.

The basic theme, adaptation and individual concerns, addresses how to protect personal assets and satisfy individual needs during the adaptation. Salehie et al. [S88] provided an approach to protecting users' assets, security requirements, from attacks. In their approach, assets are expressed in a goal model while objectives of an attacker are expressed with a threat model. A causal network is built based on the two models for deriving a set of countermeasures to mitigate threats. Omoronyia et al. [S90] proposed a framework that exploits the notion of privacy awareness requirements and supports decision making in order to minimize exposure to privacy threats and maximize functional benefits. Besides, Liaskos et al. [S89] presented a requirements-driven approach to customizing behaviors of software systems. Goal models are constructed to represent alternative behaviors and customization information is then added to restrict the space of behavioral possibilities. Custom preferences posed at the requirements level can be directly translated into system customizations through mapping elements of goal models to units of source code.

*T3.2 Verifying requirements at runtime*

The first basic theme addresses how to verify requirements based on models that specify the adaptation logic. Goldsby et al. [S91] introduced AMOEBA-RT, a runtime monitoring and verification technique that provides assurance for DAS. The verified adaptive programs are expressed with the logic proposed in [S29, S30]. Similar work can be found in Zhao et al. [S92]. They separated functional behavior from adaptation behavior of adaptive programs. The former is modeled as state machines while the latter is modeled as mode automata. Local, adaptation and global properties are specified with mode-extended LTL and verified in NuSMV model checker. For guaranteeing robustness requirements, de la Iglesia and Weyns [S93] built adaptation logic with MAPE loops. The requirements are specified with CLT and verified in UPPAAL model checker. Similar work can be found in Iftikhar and Weyns [S94].

The second basic theme focuses on verifying probabilistic requirements involved in reliability models. Filieri and Tamburrelli [S95] and Filieri et al. [S96] proposed a mathematical framework for probabilistic model checking at runtime. The reliability model is built as a DTMC and reliability requirements are expressed with PCTL. The verification process is carried out in PRISM model checker. In their further research, Filieri et al. [S97] focused on the assurance of NFR, i.e., reliability and performance, for service systems. Except for DTMC, a CTMC is adopted as the performance model at runtime. These studies provide efficient model checking techniques for supporting assurance in SASs at runtime.

The theme, verification with models@runtime, concerns how to deals with requirements verification in changing environments. Parametric models can support reasoning about design decisions and requirements verification. These models are built based on numerical estimates of domain experts. However, these parameters may change over time in the dynamic environments. To deal with this problem, Epifani et al. [S98] proposed to utilize Bayesian estimator to deal with runtime data and produce updated parameters. By verifying with the updated model, it is possible to detect or predict if a property is or will be violated.

The last basic theme identifies the verification processes. Weyns [S99] proposed the MAPE based adaptation logic and the verification process in [S93, S94]. Zhao et al. [S100] introduced the process of verifying adaptive programs in [S92].

*T3.3 Monitoring requirements at runtime*

This organizing theme deals with how to detect requirements violations through monitoring method. Though the verification activity can also reveal whether a requirement is satisfied or not, it relies on the representation of the system, the expression of verified properties and model checking techniques. In this theme, monitoring relies on the gauged contextual variables and requirements models.

For the first basic theme, Pasquale and Spoletini [S101] illustrated how to use the FLAGS language [S9] to express the requirements of service compositions and provided a technique to monitor them at runtime. Wang et al. [S102, S103] proposed to monitor satisfaction of requirements and diagnose what went wrong by modeling requirements with KAOS method and transforming the diagnostic problem into a propositional satisfaction problem that can be solved by SAT solvers. Ramirez and Cheng [S104] leveraged utility to represent the satisfaction of a requirement which can be calculated with a utility function. The monitoring problem is converted to how to calculate kinds of utilities. For the other basic theme, Ramirez et al. [S105] provided Plato-RE, an evolutionary computation based approach to monitoring the satisfaction of requirements. The approach can generate new monitoring configurations to minimize monitoring costs and maximize monitoring accuracy.

*T3.4 System evolution at runtime*

This theme considers *system evolution driven by requirements evolution*. Studies in this theme apply model-driven techniques and consider models as runtime entities. Souza et al. [S106, S107] identified evolution requirements, which specify changes to other requirements when certain conditions apply. They also modeled this type of requirement as notations to KAOS model and proposed how to operationalize them within *Zanshin* framework. Alferez and Pelechano [S108] presented a feature model based framework to support the dynamic evolution of context-aware systems, to deal with unexpected context events in the open world. When adaptation decisions are not enough to solve uncertainties, the evolution planner will guide the evolution of supporting models. Compared with [S108], Inverardi and Mori [S109] also investigated the evolution of context-aware systems from the feature engineering perspective. However, their approach does not involve the adaptation level. They provided a generic framework to deal with unforeseen evolution caused by context requirements changes.

## 4. THREATS TO VALIDITY
### 4.1 Threats to search sources

Threats to search sources might arise during choosing publication venues and digital libraries. The defined publication venues for manual search may not cover all the publications of the literature. We mitigated this threat by conducting the automated search and the "snowballing" search. We limited the manual search based on publication venues using the ERA ranking. We believe that by including in top journals, conferences and symposiums, the quality of the results of our SLR can be improved a lot. To guarantee thoroughly retrieved results, we identified eight popular digital libraries that almost cover all the publications of the literature. We did not choose Google Scholar as a digital library in our study, because the number of retrieved results is quite large and do not match well. Besides, we did not choose SCOPUS, since the retrieved results of SCOPUS are duplicated from results of ScienceDirect.

### 4.2 Threats to search strategies

Since different digital libraries provide different capabilities to search for publications, we conducted a pilot study in each digital libraries to check the validity and feasibility of the search strings. Then according to the performance of the digital libraries, we divided them into two groups and designed search strings for each group. Though the search strings can be broad, it is possible that they are not able to capture some studies. It can be mitigated through a "snowballing" search. Moreover, the time scope of automated search is from 2003 to 2013. There possibly exist several relevant studies before 2003. However, the selected results depict that no studies existed in 2003 and 2004, which may indicate the possibility is quite slim.

### 4.3 Threats to selection strategies

Threats to selection strategies might arise during defining selection criteria and selecting primary studies. The main threat to the validity of selection criteria is that we excluded editorial, abstract, keynote, poster and short paper within 6 pages or less (**C6**), opinion papers and position papers which have no detailed description of modeling methods (**C7**), and all the secondary studies (**C8**). Actually, we did not intend to exhaustively list all the publications in the literature. The objective of this study is to primarily investigate the support of modeling methods for requirements activities. Therefore, we only focus on the studies which involve more detailed descriptions of modeling methods, requirements activities and demonstrations. Meanwhile, these highly qualified studies can improve the quality of our SLR. We also conducted a pilot study for ensuring the effectiveness of the selection criteria.

The main threat to the selection process is researchers' bias. To reduce such bias, we produced a well-defined protocol to guarantee the consistency in the selection of primary studies. Before selection, we conducted a pilot study to ensure the validity of the defined selection process. The selection process was conducted in parallel by two researchers and a cross-check was performed after each selection round. Besides, we calculated Kappa coefficient to certify the inter-rater agreement between the two researchers. When any disagreement between the researchers occurred, we conducted a joint meeting with other researchers and eliminated the disagreement through discussion.

### 4.4 Threats to data extraction

Threats might also arise during the data extraction process. First, we utilized the protocol to ensure the consistency of the extraction process for two researchers. The extraction results heavily rely upon their understanding of the content of primary studies, especially modeling methods and requirements activities. To minimize the researchers' bias, we still leveraged Kappa coefficient, cross-check and joint meeting.

## 5. RELATED WORK

In this section, we present and discuss some previous secondary studies relevant to self-adaptive systems. From this perspective, we identified three studies related to our work, which focus on control engineering approaches, formal methods and research subjects, respectively. The differences between these studies and our study are presented in Table 7.

Patikirikorala et al. 1) built a classification model of the existing literature; 2) quantified the published research work on the various modeling, control schemes and validation techniques; and 3) analyzed the clustering of papers across categories of the classification model and identified research trends. However, we consider using different control schemes to engineer SASs from RE perspective. More specifically, we focus more on how to apply these control schemes to build effective adaptation mechanisms at requirements time.

In [10], Weyns et al. 1) presented research trends in applying formal methods; 2) identified the used modeling languages and property specification languages; and 3) categorized the concerned quality attributes, software systems and verified properties.

Different from their work, we categorize the modeling methods according to the nature of usage. Apart from the advantage of systematization, our SLR contains (but is not limited to) investigating formal methods and verification activities. Besides, the synthesized results and the generated conclusions of our SLR are based on more relevant studies and more recently published work.

In [11], Weyns et al. categorized the research subjects, summarized the used application domains, identified their concrete focus and claimed benefits. Different from their work, we concentrate on the RE aspect of self-adaptive systems, particularly the requirements modeling and analysis aspect. Apart from the classification of modeling methods, we also provide a systematic classification of requirements activities which is useful for researchers and practitioners to better understand the literature. Besides, we present how the identified modeling methods can support each requirements activity in each research themes.

Table 7. Comparison between related SLRs

| Related SLR | Literature | Time span | # of digital library | # of relevant venues | # of selected papers | Consider modeling methods | Consider requirements activities | Consider research themes |
|---|---|---|---|---|---|---|---|---|
| Patikirikorala et al. [8] | Control engineering for SASs | 10 years | 4 | 12 | 161 | × | × | × |
| Weyns et al. [9] | Formal method for SASs | 10 years | N/A | 16 | 75 | √ | × | × |
| Weyns et al. [10] | SASs | 6 years | N/A | 2 | 96 | × | × | × |
| Our SLR | RE of SASs | 10 years | 8 | 21 | 109 | √ | √ | √ |

# 6. CONCLUSIONS

During the past decade, an increasing number of publications in the RE community indicates the growing interest in self-adaptive systems. In this paper, we report a systematic literature review that investigates and summarizes the state of the art on requirements modeling and analysis for self-adaptive systems. We aim to identify the used modeling methods, requirements activities and research themes.

By systematically analyzing the selected 109 papers, we extracted 29 modeling methods, 14 requirements activities, 35 basic themes, 12 organizing themes and 3 global themes. By answering RQ4, we detailed how these modeling methods can support each research theme. This work will contribute to the understanding of requirements modeling and analysis for self-adaptive systems. Researchers and practitioners from both the RE field and the SASs field will benefit from this work.

Future work focuses on further investigating the relationship between requirements modeling methods and RE activities. We plan to carry out a literature review of the assurance method for requirements-driven adaptation. Besides, we are also interested in building adaptation mechanisms to deal with runtime uncertainties in the relevant research themes.

# 7. ACKNOWLEDGMENTS

We thank Prof. Barbara A. Kitchenham and her team at Keele University for reviewing our protocol and all the received advice. This research is supported by the National Natural Science Foundation of China under Grant Nos 61232015 and 91318301, and in part by the Natural Science Foundation of Guangxi Province under Grant No. 2012GXNSFCA053010.

## APPENDIX A. SELECTED STUDIES